\documentclass[lettersize,journal]{IEEEtran}
\usepackage{amsmath,amsfonts}
\usepackage{balance}
\usepackage{color}
\usepackage{algorithmic}
\usepackage{algorithm}
\usepackage{array}
\usepackage[caption=false,font=normalsize,labelfont=sf,textfont=sf]{subfig}
\usepackage{textcomp}
\usepackage{stfloats}
\usepackage{url}
\usepackage{verbatim}
\usepackage{graphicx}
\usepackage{cite}
\begin{document}
	
	\title{Backscatter-Assisted High-Speed Rail Communications in Straight Tunnel Environments: Effects of Tag Number and Phase Control
	}

	\author{Yunping~Mu,
		Gongpu~Wang,~\IEEEmembership{Member, IEEE},
		Ruisi~He,~\IEEEmembership{Senior Member, IEEE},
		Theodoros A. Tsiftsis,~\IEEEmembership{Senior Member,~IEEE},
		Saman~Atapattu,~\IEEEmembership{Senior Member, IEEE}, and
		Chintha~Tellambura,~\IEEEmembership{Fellow, IEEE}
		
		\thanks{This work was supported by Beijing Municipal Natural Science Foundation (No.~L224008) and the Natural Science Foundation of China (No.~U22B2004). The work of T. A. Tsiftsis was supported by the project NEURONAS. The research project NEURONAS is implemented in the framework of H.F.R.I call ``3rd Call for H.F.R.I.’s Research Projects to Support Faculty Members \& Researchers'' (H.F.R.I. Project Number: 25726).}
        
		\thanks{Y. Mu and G. Wang are with Engineering Research Center of Network Management Technology for High Speed Railway of Ministry of Education, School of Computer and Information Technology, Beijing Jiaotong University, Beijing 100044, China. Email:myp0530@163.com, gpwang@bjtu.edu.cn.}
				
		\thanks{R. He is with the State Key Laboratory of Rail Traffic Control and Safety, Beijing Jiaotong University, Beijing 100044, China. Email: ruisi.he@ieee.org.}
		
		\thanks{T. A. Tsiftsis is with the Department of Informatics and Telecommunications, University of Thessaly, Lamia 35100, Greece. Email: tsiftsis@uth.gr.}
		
		\thanks{S. Atapattu is with the School of Engineering, RMIT University, Melbourne, Victoria, Australia. Email: saman.atapattu@rmit.edu.au.}
		
		\thanks{C. Tellambura is with the Department of Electrical and Computer Engineering, University of Alberta, Edmonton, AB T6G 1H9, Canada. Email:chintha@ece.ualberta.ca.}
				
	}

	\maketitle

	\begin{abstract}
		Backscatter communication is a promising technology to enhance the signal strength received by the receiver in straight tunnel environments. The impact of the number of tags and their phase adjustment on system performance remains a challenging issue though. Therefore, in this paper, we investigate the channel gain of backscatter-assisted communication with multiple tags in straight tunnels. In particular, we derive the probabilities that the backscatter link gain is greater than the direct link under adjustable and random phase assumptions by applying the Gaussian and Gamma approximations to derive tractable expressions.
		The simulation results show that phase-adjustable tags significantly improve the channel gain of the backscatter links compared to the random phase case. Moreover, the number of tags has an upper threshold for an effective tag deployment pattern.
		These insights provide valuable guidelines for the efficient design of backscatter communication systems in tunnel environments.
	\end{abstract}
		

	\IEEEpeerreviewmaketitle

	\section{Introduction}
	\IEEEPARstart{H}{igh-speed} rail (HSR) has increasingly emerged as the favored mode of transportation, driven by its high speed and convenience. This shift also heightens the demand for wireless information services among passengers.

	Notably, tunnels constitute a significant and growing proportion of HSR environments in recent years. However, wireless communication within these tunnels faces three primary challenges.      
	First, the operating speed of trains on   HSR can reach $350$ km/h, while the coverage radius of base stations is around $1$ km. Therefore, rapid movement can cause the Doppler effect and frequent cell switching, which reduces communication quality. 
	Next, the narrow and enclosed space inside the tunnel as well as the rough inner wall leads to more reflection, diffraction, and scattering of wireless signals, resulting in severe signal attenuation. In addition, one HSR carriage has $60-100$ people, and the demand data rate for passengers in each carriage is about $37.5$ Mbps. The above factors pose crucial challenges to wireless communication in HSR tunnel scenarios \cite{Radio Wave Propagation Scene Partitioning for High-Speed Rails_Int J Antenn Propag_2016, 3D Non-Stationary Wideband Tunnel Channel Models for 5G High-Speed Train Wireless Communications_IEEE T Intell Transp_2020, Intelligent Beam Management Based on Deep Reinforcement Learning_IEEE Trans. Veh. Tech._2024}.

	Current wireless coverage solutions in tunnels include leaky coaxial cables, leaky waveguides, and specific type of antennas. Among these, customized antennas can be installed at the ends of either short tunnels or within long tunnels \cite{DMRS-based channel estimation for railway communications in tunnel environments_Veh Commun._2021}.  Benchmarked against the other two methods, special antennas exhibit notable advantages, including straightforward installation, effortless maintenance, and reduced capital expenditure, thereby serving as the primary focus of this paper.

	Backscatter communication has emerged as a transformative technology in the Internet of Things (IoT) landscape \cite{Breaking the Interference and Fading Gridlock in Backscatter Communications_IEEE Commun._2024,Activation Distance and Capacity Analysis for Ambient Backscatter Communications with Sensitivity Constraint and Beamforming_China Commun._2023}, offering battery-free operation and significant cost reduction for wireless sensors. Its integration with cellular networks has garnered significant academic interest due to its distinctive advantages, including zero power consumption, minimal hardware requirements, low maintenance overhead, ease of deployment, and flexible extension \cite{Channel Estimation for Backscatter Communication Systems_IET Commu_2024,Wavy Signals and Striped Constellations for Backscatter Communications_IEEE Trans. Wireless Commun._2024}. Notably, backscatter technology was incorporated into HSR named as backscatter-assisted wireless transmission (BAWT) \cite{Backscatter Aided Wireless Communications on High Speed Rails: Capacity Analysis and Transceiver Design_IEEE J. Sel. Area. Comm._2020}.

	In \cite{Backscatter Assisted Wireless Communications_China Commun_2024}, the backscatter technology to enhance the coverage of special antennas in straight tunnels has been implemented. However, the auxiliary effect of a single tag remains restricted, necessitating the deployment of multiple tags for superior performance. The number of tags, and whether the tag phase is adjustable become critical, which motivates our current study. In this paper, we investigate the probability that the channel gain of backscatter link exceeds that of the direct link. This metric is approximately studied by using the Gaussian and Gamma probability density functions (PDFs). Based on the above metric, an appropriate choice of tag number and phase control strategy can be selected to ensure high quality communication links.

	\section{System Model}\label{sec:SysModel}
	
\begin{figure}[t!]
		\centering
		\includegraphics[width=2in]{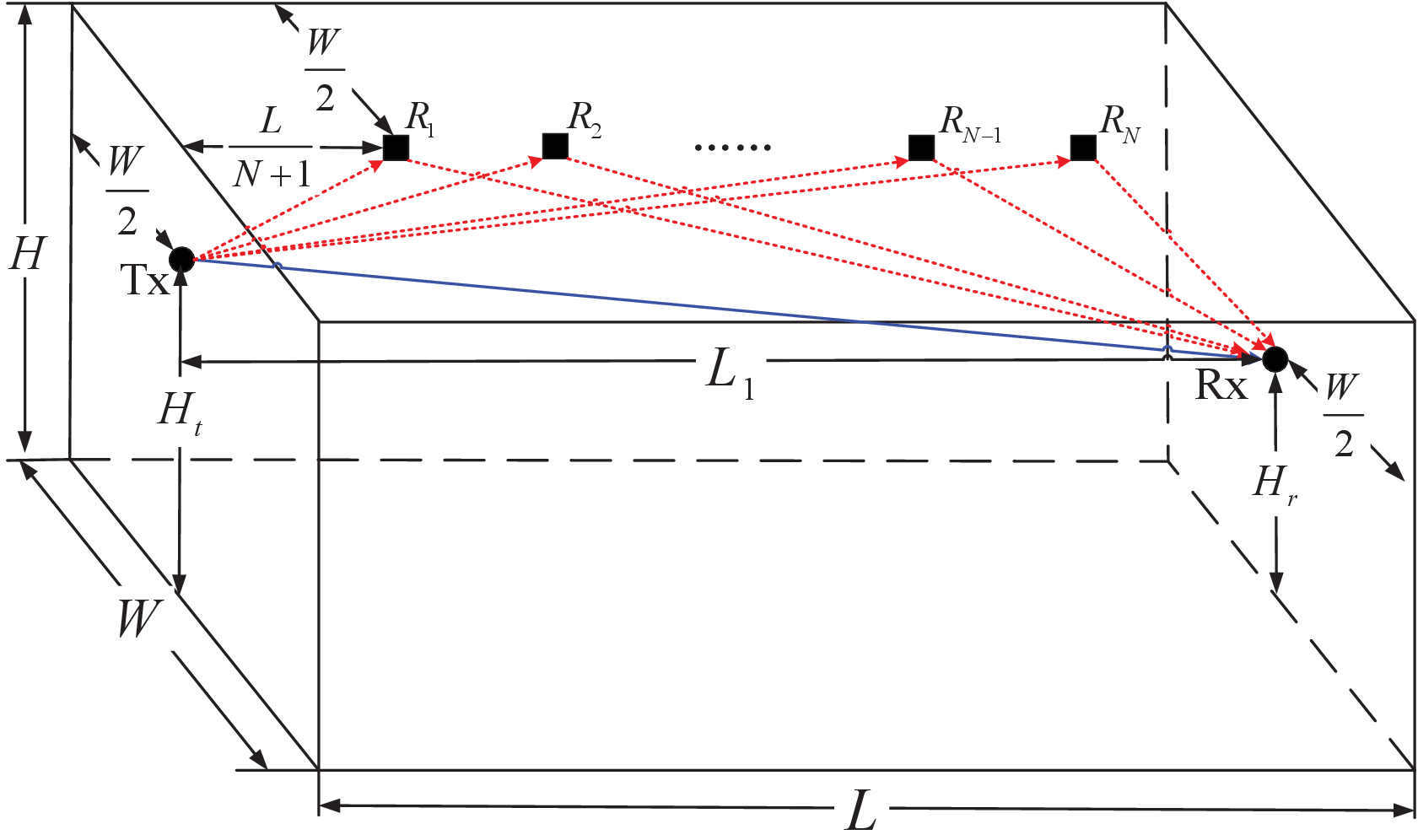}
		\caption{Backscatter-assisted  wireless communication system for straight tunnel scenario.}
		%: $N$ backscatter tags are evenly arranged in a row on the ceiling of the tunnel%}}
		\label{fig_1_Tag assisted tunnel scenarios}
	\end{figure}

	In this paper, we consider the deployment of $N$ tags on the ceiling of the tunnel, as illustrated in Fig. \ref{fig_1_Tag assisted tunnel scenarios}. The transmit antenna  (Tx) is installed in the center of the tunnel entrance with a distance $H_t$ from the ground, while the receive antenna (Rx) is  mounted on the top of the train at a height of $H_r$ from the ground \cite{Two-Slope Path Loss Model for Curved-Tunnel Environment With Concept of Break Point_IEEE Trans. Intell. Transp._2021, Measurement of Distributed Antenna Systems at 2.4 GHz in a Realistic Subway Tunnel Environment_IEEE Trans. Veh. Technol._2012}. Notably, $H_t>H_r$, with the tunnel dimensions given by height $H$, length $L$, and width $W$. The horizontal distance between Tx and Rx is $L_1(n)$. To achieve better coverage and seamless connectivity, we distribute tags throughout the tunnel, installing them on the ceiling along the direction of movement of the train. 
    In modeling, we consider the direct channel between the Tx and the Rx and the single-bounce backscatter channel through each tag\footnote{The selected channel modeling is supported by measured propagation characteristics in tunnel environments, where the LoS component typically contributes more than half of the received power, and dominant single-bounce reflections account for an additional 20-30\%. These two components together represent over 90\% of the extracted path energy \cite{Wideband Polarimetric Directional Propagation Channel Analysis Inside an Arched Tunnel_IEEE Trans. Antenn. Propag._2009}. Moreover, the spatio-temporal analysis in \cite{3D Non-Stationary Wideband Tunnel Channel Models for 5G High-Speed Train Wireless Communications_IEEE T Intell Transp_2020} indicates that the temporal correlation characteristics of the channel are primarily determined by the LoS and first-order reflections, while higher-order scattering mainly forms low-power diffuse components. Therefore, focusing on the LoS and single-bounce terms captures the dominant propagation mechanisms and preserves physical realism while keeping the model sufficiently tractable for performance analysis.}. The tags do not mutually interact and the wireless channel exhibits time-varying characteristics\footnote{The inter-tag reflection paths undergo double-bounce attenuation and are therefore negligible compared to the single-bounce backscatter link. Furthermore, the physical spacing between adjacent tags along the tunnel ceiling is on the order of $\lambda/2$ or larger under the considered deployment, and the high-speed motion of the train induces rapid temporal phase decorrelation. These factors jointly ensure that the backscatter channels corresponding to different tags are statistically independent \cite{Transmissive Metasurfaces Assisted Wireless Communications on Railways: Channel Strength Evaluation and Performance Analysis_IEEE Trans. Commun._2023}.}.   
    
	Let $h(t)$, ${f_i(t)}$, and $g_i(t)$ denote the channel coefficients from the Tx to the Rx, from the Tx to the $i$th ($i \in \left\{ {1,2,...,N} \right\} $) tag, and from the $i$th tag to the Rx, respectively, and the corresponding distances are $d_h(t)$, $d_{fi}(t)$, and $d_{gi}(t)$, respectively. Here, $d_h^2(t) = \left( H_t-H_r \right)^2 + L_1^2(t)$,
	$d_{fi}^2(t) = (H-H_t)^2 + {(\frac{{iL}}{{N + 1}})^2}$, and $d_{gi}^2(t) = {\left( {H - {H_r}} \right)^2} + ({L_1(t) - \frac{{iL}}{{N + 1}})^2}$.  
	All small-scale fading channel coefficients are assumed to be independent and identically distributed zero-mean and unit-variance complex Gaussian variables, i.e., $h(t),f_{i}(t), g_{i}(t) \sim \mathcal{CN}(0,1)$. 
	
	In this system, the Tx transmits the signal $s(n)$ with carrier frequency $f_1,$ and initial phase $\theta_1$. The signals received by the Rx  can be formulated as
	\begin{align}
	\tilde y(t) &= \eta \left( {\sum\limits_{i = 1}^N {\frac{{{g_i}(t){f_i}(t){e^{j\left( {2\pi f_D^{\left( i \right)}t + {\phi _i}} \right)}}}}{{\sqrt {d_{{g_i}}^\alpha (t)d_{{f_i}}^\alpha (t)} }}} } \right)s(t)x(t) {e^{j\left( {2\pi {f_1}t + {\theta _1}} \right)}}\notag\\ 
	& \qquad + \frac{{{h}(t){e^{j2\pi {f_d}t}}}}{{\sqrt {d_h^\alpha (t)} }}s(t)  {e^{j\left( {2\pi {f_1}t + {\theta _1}} \right)}} + {w_1}(t),
	\label{eq:y_1}
	\end{align}
	where $\eta \in (0,1]$ is the attenuation factor of passive tag,  $\alpha$ is the large-scale path loss factor, $x(t) \in \{ -1,1\}$ is the binary data signals of the passive tag, ${w_1}(t)$  represents zero-mean complex additive white  Gaussian noise with variances $\sigma_{w1}^2$, $\phi_i$ is the phase shift introduced by the $i$th tag, ${f_D^{\left( i \right)}}$ and ${{f_d}}$ are the Doppler shifts of the $i$-th backscatter channel and the direct channel, respectively.

	Define the local carrier signal generated by the Rx with carrier frequency $f_2$ and initial phase $\theta_2$.
	Ideally, these two quantities are identical to $f_1$ and $\theta_1$, respectively.
	However, there often exists a certain carrier frequency offset (CFO) and phase error in practical applications. Thus, the Rx will demodulate the received signal $y_1(t)$ to the baseband signal as follows
	\begin{align}	
	y(t) &= \tilde y(t){e^{ - j\left( {2\pi {f_2}t + {\theta _2}} \right)}}\notag\\
	&=\eta \left( {\sum\limits_{i = 1}^N {\frac{{{g_i}(t){f_i}(t){e^{j\left( {2\pi f_D^{\left( i \right)}t + {\phi _i}} \right)}}}}{{\sqrt {d_{{g_i}}^\alpha (t)d_{{f_i}}^\alpha (t)} }}} } \right)s(t)x(t){e^{j\left( {2\pi {\Delta _f}t + {\Delta _\theta }} \right)}}\notag\\
	&\qquad+ \frac{{{h}(t){e^{j2\pi {f_d}t}}}}{{\sqrt {d_h^\alpha (t)} }}s(t){e^{j\left( {2\pi {\Delta _f}t + {\Delta _\theta }} \right)}} + {w_2}(t),
	\label{eq:y_nt}
	\end{align}
	where $\Delta_f=f_1-f_2$ represents the CFO, $\Delta_\theta=\theta_1-\theta_2$ denotes the phase error, and ${w_2}(t)$ represent zero mean additive complex  Gaussian noise with variance $\sigma_{w2}^2$.
	
	Assume that the sampling period $T_s=\frac{{{1}}}{{{f_s}}}$ and the sampling point $t=nT_s$, then we can obtain 
	\begin{align}
	& y(n) = y(t)\left| {_{t = n{T_s}}} \right.\notag\\
	&\quad= \eta \left( {\sum\limits_{i = 1}^N {\frac{{{g_i}(n){f_i}(n){e^{j\left( {2\pi \varepsilon _D^{\left( i \right)}n + {\phi _i}} \right)}}}}{{\sqrt {d_{{g_i}}^\alpha ({n})d_{{f_i}}^\alpha (n)} }}} } \right)s(n)x(n){e^{j\left( {2\pi {\varepsilon _f}n + {\Delta _\theta }} \right)}}\notag\\
	&\quad+ \frac{{{h}(n){e^{j2\pi {\varepsilon _d}n}}}}{{\sqrt {d_h^\alpha (n)} }}s(n){e^{j\left( {2\pi {\varepsilon _f}n + {\Delta _\theta }} \right)}} + {w_2}(n),
	\label{eq:y_n}
	\end{align}
	where $\varepsilon _D^{\left( i \right)} = \frac{{f_D^{\left( i \right)}}}{{{f_s}}}, {\varepsilon _d} = \frac{{{f_d}}}{{{f_s}}},  {\varepsilon _f} = \frac{{{\Delta _f}}}{{{f_s}}}$ are the normalized Doppler shift of the $i$-th backscatter channel, the normalized Doppler shift of the direct channel, and the normalized CFO, respectively.
	
	\section{Performance Analysis}
	\label{sec:performance analysis}
	
	First we assume pilot-aided estimation and compensation of dominant CFO and Doppler shift \cite{Channel Estimation and Pilot Design for Ambient Internet of Things With Frequency Offsets_IEEE Trans. Veh. Tech._2025, Doppler mitigation method aided by reconfigurable intelligent surfaces for high-speed channels_WCL_2022}. Then, we explicitly model the remaining residual phase as a sum of a common component $\Psi(n)$ and residual perturbation in each backscatter link $\delta_i(n)$. Consequently, (\ref{eq:y_n}) can be rewritten as
    \begin{align}
      y(n) &= \eta \sum_{i=1}^N \frac{g_i(n) f_i(n) s(n) x(n) e^{j(\Psi(n)+\phi_i
      + \delta_i(n))}  }{\sqrt{d_{g_i}^\alpha(n) d_{f_i}^\alpha(n)}}   \notag\\
	   &+ \frac{h(n) s(n) e^{j\Psi(n)}}{\sqrt{d_h^\alpha(n)}}+ w_2(n),
    \label{eq:y_simplified}
    \end{align}
    where $\phi_i$ can be intentionally controlled or random.

	We define the metric
	\begin{align}
	P(N) = \Pr \left( |\eta A_r|^2 \ge |A_0|^2 \right),\, i=1,\dots,N
	\label{eq:prob_def}
	\end{align}
	where the index $n$ in all parameters is omitted for the sake of simplicity, and $ A_r \mathop = \limits^\Delta  e^{j\Psi}\sum_{i=1}^N \frac{g_if_i e^{j(\phi_i+\delta_i)}}{\sqrt{d_{g_i}^\alpha d_{f_i}^\alpha}}, A_0\mathop = \limits^\Delta  \frac{he^{j\Psi}}{\sqrt{d_h^\alpha}}$
	under two different phase assumptions:
	\begin{itemize}
		\item \textbf{Adjustable tag phases:} $\phi_i$ can be set (e.g., via intelligent backscatter control) to maximize the coherent sum. 
        %to coherently align the backscatter signals, achieving constructive combination at the Rx.
		\item \textbf{Random tag phases:} $\phi_i \stackrel{\text{i.i.d.}}{\sim}\mathcal{U}[0,2\pi)$.
	\end{itemize}
	Since the common factor \(e^{j\Psi}\) cancels out in the inequality of \eqref{eq:prob_def} and therefore does not affect the statistics of the decision metric, we use $ A_r \mathop = \limits^\Delta  \sum_{i=1}^N \frac{g_if_i e^{j(\phi_i+\delta_i)}}{\sqrt{d_{g_i}^\alpha d_{f_i}^\alpha}}, A_0\mathop = \limits^\Delta  \frac{h}{\sqrt{d_h^\alpha}}$ in the following analysis.

	\subsection{ Adjustable Tag Phases}
	\label{sec: Adjustable Tag Phases}
   In this subsection, we first consider an idealized benchmark in which the effective phase of each backscatter channel can be estimated within
   the channel coherence time and the tag phases are accordingly adjusted for coherent combining. In this case,
    the optimal phase control is
    $ \phi_i = -\arg\big(g_i f_i\big) (\text{mod }2\pi),$
    so that all backscatter components add constructively at the Rx. This scenario represents the maximum achievable
    gain and thus serves as a performance upper bound for comparison.

	Since $\Pr\big(a^2|z_1|^2 \ge b^2|z_2|^2\big)= \Pr\big(a|z_1| \ge b|z_2|\big)$ when $z_1,z_2\in\mathbb{C}$ and $a,b\in\mathbb{R}^+$. Applying optimal phase control, we further rewrite (\ref{eq:prob_def}) as 
	\begin{align}
	P(N)=\Pr\!\Big(\eta \left| {\sum_{i=1}^N \frac{|g_i||f_i|}{\sqrt{d_{g_i}^\alpha d_{f_i}^\alpha}} e^{j\delta_i}} \right|\ge \frac{|h|}{{\sqrt {d_h^\alpha } }}\Big).
    % P(N)=\Pr \left( {\eta \sum\limits_{i = 1}^N {\frac{{\left| {{g_i}} \right|\left| {{f_i}} \right|}}{{\sqrt {d_{{g_i}}^\alpha {\mkern 1mu} d_{{f_i}}^\alpha } }}}  \ge \frac{{\left| {{h}} \right|}}{{\sqrt {d_h^\alpha } }}} \right)
	\label{P_final}
	\end{align}

	By defining variables 
    $G_a \mathop  = \limits^\Delta \sum\limits_{i = 1}^N a_i e^{j\delta_i}, a_i\mathop  = \limits^\Delta{\frac{{\left| {{g_i}} \right|\left| {{f_i}} \right|}}{c_i}}$, $c_i\mathop  = \limits^\Delta{\sqrt {d_{{g_i}}^\alpha {\mkern 1mu} d_{{f_i}}^\alpha } }$ and $G_b  \mathop= \limits^\Delta \left| {{h}} \right|$, we can simplify (\ref{P_final}) as
	\begin{align}
	P(N)=\Pr \left( {{G_b} \le c_0 {|G_a|}} \right),
	\label{P_N}
	\end{align}
	where $c_0=d_h^{\frac{\alpha }{2}}\eta$.
	Then, we can calculate (\ref{P_N}) as 
	$P(N)~=~ \int_0^\infty  {{f_{|Ga|}}(y)} \int_0^{c_0y} {{f_{Gb}}} (x)dxdy,$
	where $f_{|G_a|}(\cdot)$ is the PDF of $|G_a|$, $f_{G_b}(\cdot)$ is the PDF of $G_b$, and $P(N)$ holds because $|G_a|$ and $G_b$  are independent.
	Since $G_b$ is Rayleigh with PDF $f_{G_b}(x)=2x e^{-x^2},\ x\ge0$,  $P(N)$ can straightforwardly be rewritten as
	\begin{align}
	P(N) = 1 - \int_0^\infty  {{f_{|Ga|}}(y)} {e^{ - {c}{y^2}}}dy =1-\mathbb{E}\!\big[ e^{-c |G_a|^2}\big],
	\label{eq:PN_expect}
	\end{align}
	where $c=c_0^2=d_h^\alpha \eta^2.$

    The exact evaluation of \eqref{eq:PN_expect} requires the PDF of the envelope $|G_a|$, which can be expressed via the characteristic function of the complex sum. Using the Gaussian Fourier identity and taking the expectation of $|G_a|$, we obtain the representation
    \begin{align}
    P(N)=1-\frac{1}{\sqrt{4\pi c}}\int_{-\infty}^{\infty} e^{-t^2/(4c)}\Phi_{|G_a|}(t)dt,
    \label{P(N)_residual}
    \end{align}
    where $\Phi_{|G_a|}(\cdot)$ denotes the characteristic function of $|G_a|$. This one-dimensional integral is amenable to numerical quadrature but admits no simpler closed-form in elementary functions for general $N$ and arbitrary residual phases. Therefore, we adopt tractable approximations (Gaussian and Gamma approximations) to get the result of $P(N)$.
	
	\subsubsection{Gaussian  approximation}
	For moderate or large $N$, the sum $G_a \mathop  = \limits^\Delta \sum\limits_{i = 1}^N a_i e^{j\delta_i}$ can be approximated by a circular complex Gaussian random variable \(G_a\!\sim\!\mathcal{CN}(\mu_G,\sigma_G^2)\) with
	\begin{align}
	\mu_G &\mathop  = \limits^\Delta\mathbb{E}\big[Ga\big]= \sum_{i=1}^N \mathbb{E}\big[a_i e^{j\delta_i}\big] \;=\; e^{-\frac{1}{2}\sigma_\delta^2}\sum_{i=1}^N\mathbb{E}[a_i],\\
	\sigma_G^2 &\mathop  = \limits^\Delta\mathbb{E}\big[|Ga-\mu_G|^2\big] = \sum_{i=1}^N \Big(\mathbb{E}[a_i^2] - \big|\mathbb{E}[a_i e^{j\delta_i}]\big|^2\Big),
	\label{eq:mu_var_def}
	\end{align}
	where the residual perturbation $\delta_i$ is modeled as a zero-mean Gaussian random variable, i.e., $\delta_i\sim\mathcal{N}(0,\sigma_\delta^2)$, and $\delta_i$ is independent of $a_i$.
	We derive that the first and second-order origin moments of $a_i$ as
	\begin{align}
	\mathbb{E}[a_i] &= \frac{\mathbb{E}[|g_i|]\mathbb{E}[|f_i|]}{c_i}
	= \frac{\pi}{4 c_i}, \label{eq:mu_i}\\
	\mathbb{E}[a_i^2] &= \frac{\mathbb{E}[|g_i|^2]\mathbb{E}[|f_i|^2]}{c_i^2}
	= \frac{1}{c_i^2},
	\label{eq:var_ai}
	\end{align}
	Thus $Var(a_i) =\frac{1}{c_i^2} - \frac{\pi^2}{16 c_i^2}$.
    Then we get
	\begin{align}
	\mu_G = e^{-\frac{1}{2}\sigma_\delta^2}\sum_{i=1}^N\frac{\pi}{4 c_i}, \quad \sigma_G^2 = \sum_{i=1}^N \Big(\frac{1}{c_i^2} - e^{-\sigma_\delta^2}\frac{\pi^2}{16 c_i^2}\Big).
	\label{mu_i,nu_i}
	\end{align}
	Under this approximation, the random variable $|G_a|^2$ is non-central chi-square with two degrees of freedom, and the moment generating identity \eqref{eq:PN_expect} yields a closed-form:
	\begin{align}
	\mathbb{E}\big[e^{-c |G_a|^2}\big] \;=\; \frac{1}{\sqrt{1+c\sigma_G^2}}
	\exp\!\Big(-\frac{c|\mu_G|^2}{1+c\sigma_G^2}\Big). 
	\end{align}
	Therefore the Gaussian approximation yields the simple closed form
	\begin{align}
	\,P_{\mathrm{Gauss}}(N) \approx 1 - \frac{1}{\sqrt{1+c\sigma_G^2}}
	\exp\!\Big(-\frac{c|\mu_G|^2}{1+c\sigma_G^2}\Big). 
	\label{eq:PN_gauss_final}
	\end{align}

	\subsubsection{Gamma approximation}
	Since the Gaussian distribution is more suitable for the case where $N$ is relatively large, the Gamma distribution is more appropriate when the value of $N$ is small.  Define $Z\mathop  = \limits^\Delta |G_a|^2$. The first two moments of $Z$ are
    \begin{align}
    \mathbb{E}[Z] = |\mu_G|^2 + \sigma_G^2,\quad
    \mathrm{Var}(Z)=2\sigma_G^4+4\sigma_G^2|\mu_G|^2.
    \end{align}
    We assume that $Z$ follows a $\mathrm{Gamma}(k,\theta)$ distribution.   Using moment matching, we get 
	\begin{align}
	k = \frac{\mathbb{E}[Z]^2}{\mathrm{Var}(Z)},\quad
    \theta = \frac{\mathrm{Var}(Z)}{\mathbb{E}[Z]}.
	\label{eq:gamma parameter1}
	\end{align}
    Using this match and the Moment Generating Function (MGF), we can calculate
	\begin{align}
	\mathbb{E}[e^{-c|G_a|^2}] = (1+c\theta)^{-k}.
	\label{eq:E_gamma_int}
	\end{align}
	 Hence the Gamma approximation gives
	\begin{align}
	P_{\mathrm{Gamma}}(N)
    \approx
    1 - \left(1 + c\theta\right)^{-k}.
	\label{eq:PN_gamma_final}
	\end{align}
	\subsection{ Random Tag Phases}
	\label{sec: Random Tag Phases}
	Consider the case where tag phases are uncontrolled, we know that the complex random variables $a_i e^{j(\phi_i+\delta_i)}$ have phases $\phi_i+\delta_i$. The tag phases $\phi_i \stackrel{\text{i.i.d.}}{\sim}\mathcal{U}[0,2\pi)$ and independent of the residuals $\delta_i$.
	Recall the metric in \eqref{eq:prob_def} and define the power sum 
	$T \mathop  = \limits^\Delta \sum_{i=1}^N a_i^2 
	= \sum_{i=1}^N \frac{|g_i|^2\,|f_i|^2}{d_{g_i}^\alpha\,d_{f_i}^\alpha},$
	where $a_i$ is defined the same as in \ref{sec: Adjustable Tag Phases}.
	Conditioned on $a_i$ (equivalently on $T$),  the in-phase and quadrature components of the complex sum $\sum_{i=1}^N a_i e^{j(\phi_i+\delta_i)}$ are zero-mean and conditionally Gaussian with variance $\tfrac{1}{2}T$. Thus the backscatter envelope is conditionally Rayleigh, i.e, 
	$|A_r|\,\big|\,T \sim \mathrm{Rayleigh}\!\left(\sigma_T\right), \sigma_T^2 = \frac{1}{2}T.$
	Consequently,
	$ |A_r|^2 \,\big|\, T \sim \mathrm{Exp}\!\big(\frac{1}{T}\big)$
	with mean $T$.
	
	For the direct link, we have $|h|^2\sim \mathrm{Exp}(1)$ and thus
	$|A_0|^2 ~= ~\frac{|h|^2}{d_h^\alpha} \sim \mathrm{Exp}\!\big(\frac{1}{m_0}\big),$
	with mean $m_0 \mathop  = \limits^\Delta \frac{1}{d_h^\alpha}$.
	Let $U\mathop  = \limits^\Delta |A_0|^2$ and $V\mathop  = \limits^\Delta |A_r|^2$. Therefore, the desired probability can be written as
	\begin{align}
	P(N) 
	&= \Pr\!\left(\eta^2 V > U\right) \notag\\
	&= \int_0^\infty   \int_0^\infty  {\left( {1 - {e^{ - d_h^\alpha {\eta ^2}v}}} \right)} \frac{1}{t}{e^{ - v/t}}{\mkern 1mu} dv{\mkern 1mu} {f_T}(t){\mkern 1mu} dt.\notag\\
	&=1-\mathbb{E}_T\!\left[\frac{1}{1 + c_0^2 T}\right],
	\label{eq:P_random_exact_expect}
	\end{align}
	where $c_0=d_h^{\frac{\alpha }{2}}\eta$.
    Importantly, the phases $\phi_i$ and the residual perturbation $\delta_i$ do not appear explicitly in \eqref{eq:P_random_exact_expect} because they have been averaged out when deriving the marginal distribution of the backscatter power $V$.
	
	To make \eqref{eq:P_random_exact_expect} explicit, write $T=\sum_{i=1}^{N} Z_i$ with
	$ Z_i ~\mathop  = \limits^\Delta ~\frac{|g_i|^2\,|f_i|^2}{d_{g_i}^\alpha d_{f_i}^\alpha}
	= w_i\, X_i Y_i,$
	where $w_i ~\mathop  = \limits^\Delta ~\frac{1}{d_{g_i}^\alpha d_{f_i}^\alpha},
	X_i=|g_i|^2, Y_i=|f_i|^2 \stackrel{\text{i.i.d.}}{\sim} \mathrm{Exp}(1).$

	Using the identity
	$ \frac{1}{1+x} = \int_{0}^{\infty} e^{-u} e^{-u x}\,\mathrm{d}u,$
	we obtain
	\begin{align}
	\mathbb{E}_T\!\left[\frac{1}{1 + c_0^2 T}\right]
	= \int_{0}^{\infty} e^{-u}\,\underbrace{\mathcal{L}_T(c_0^2 u)}_{\prod_{i=1}^{N}\mathcal{L}_{Z_i}(c_0^2 u)}\,\mathrm{d}u,
	\label{eq:LT_T}
	\end{align}
	where $\mathcal{L}_T(s)=\mathbb{E}[e^{-sT}]$ is the Laplace transform (LT) of $T$. For each $Z_i=w_i X_i Y_i$,
	\begin{align}
	\mathcal{L}_{Z_i}(s) 
	&= \mathbb{E}\!\left[\frac{1}{1 + s w_i Y}\right] 
	= \frac{e^{\,\frac{1}{s w_i}}}{s w_i}\, E_1\!\left(\frac{1}{s w_i}\right),
	\label{eq:LZi}
	\end{align}
	where $E_1(x)=\int_x^{\infty} \frac{e^{-t}}{t}\,\mathrm{d}t$ is the exponential integral. 
	
	Combining \eqref{eq:P_random_exact_expect}-\eqref{eq:LZi},  we derive
	\begin{align}
	\hspace{-0.2cm}P(N) 
	= 1 - \int_{0}^{\infty} e^{-u}\,
	\prod_{i=1}^{N} 
	\left[
	\frac{e^{\,\frac{1}{c_0^2 u\, w_i}}}{c_0^2 u\, w_i}\,
	E_1\!\left(\frac{1}{c_0^2 u\, w_i}\right)
	\right]
	\,\mathrm{d}u.
    \label{eq:P_random_exact_LT}
	\end{align}

	Similar to \eqref{P(N)_residual}, we also use two approximations to obtain the closed-form solution of $P(N)$. 
	\subsubsection{Gaussian approximation}Obviously, we have 
	$\mathbb{E}[X_i Y_i] = 1,\quad
	\mathrm{Var}(X_i Y_i) 
	= \mathbb{E}[X_i^2]\mathbb{E}[Y_i^2]-\mathbb{E}[X_i Y_i]^2
	=3.$
	Hence
	$\mu_T \mathop  = \limits^\Delta \mathbb{E}[T] = \sum_{i=1}^{N} w_i,\quad
	\sigma_T^2 \mathop  = \limits^\Delta \mathrm{Var}(T) = 3\sum_{i=1}^{N} w_i^2.$
	Since $g(t)\mathop  = \limits^\Delta \frac{c_0^2 t}{1+c_0^2 t}$ is smooth and concave, a second-order Delta approximation around $\mu_T$ gives
	\begin{align}
	P_{Gauss}(N) 
	& = \mathbb{E}[g(T)]
	\approx g(\mu_T) + \frac{1}{2}g''(\mu_T)\sigma_T^2 \notag\\
	&= \frac{c_0^2 \mu_T}{1+c_0^2 \mu_T}
	\;-\; \frac{c_0^4 \sigma_T^2}{(1+c_0^2 \mu_T)^3}.
	\label{eq:P_gauss_delta}
	\end{align}
	\subsubsection{Gamma approximation} Approximate $T$ by a Gamma random variable $T_{\Gamma}\sim \mathrm{Gamma}(k_T,\theta_T)$ whose mean and variance match $\mu_T\,,\sigma_T^2$:
	${k_T} = {{\mu _T^2} \mathord{\left/ {\vphantom {{\mu _T^2} {\sigma _T^2}}} \right. \kern-\nulldelimiterspace} {\sigma _T^2}},{\theta _T} = {{\sigma _T^2} \mathord{\left/ {\vphantom {{\sigma _T^2} {{\mu _T}}}} \right.\kern-\nulldelimiterspace} {{\mu _T}}}.$
	Using the identity 
	$\int_{0}^{\infty} e^{-u}(1+\theta_T c_0^2 u)^{-k_T}\mathrm{d}u 
	= e^{\frac{1}{\theta_T c_0^2}} (\theta_T c_0^2)^{k_T-1}\,\Gamma\!\left(1-k_T,\,\frac{1}{\theta_T c_0^2}\right),$
	we obtain
	\begin{align}
	\hspace{-0.2cm} P_{Gamma}(N)
	= 1 - e^{\frac{1}{\theta_T c_0^2}}\,(\theta_T c_0^2)^{\,k_T-1}\,
	\Gamma\!\left(1-k_T,\,\frac{1}{\theta_T c_0^2}\right),
	\label{eq:P_gamma_random}
	\end{align}
	where  $\Gamma(\cdot,\cdot)$ is the upper incomplete Gamma function. 
	\begin{figure}[ht]
		\centering		\includegraphics[width=2in]{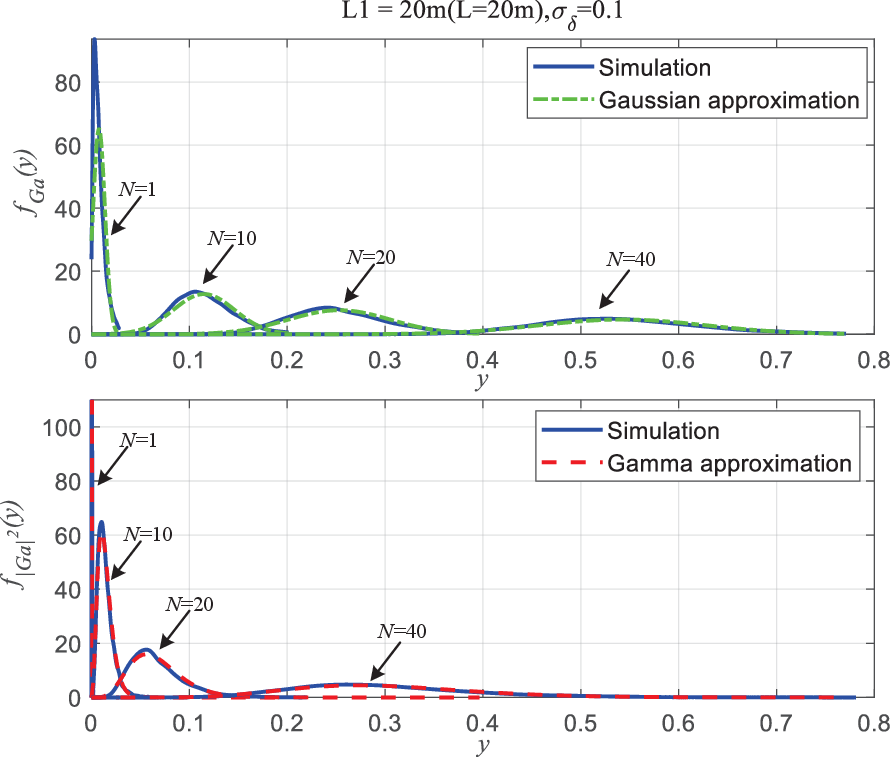}
		\caption{PDFs of $G_a$ and  $|G_a|^2$.}
		\label{fig_2_PDF}
	\end{figure}
	\section{simulation results}
	\label{sec:simulation results}
	This section provides simulation results aimed at validating the results derived in Section \ref{sec:performance analysis}.
	We set $\eta=0.5$, $\alpha=2$, $H=5.3$ m, $W=4.8$ m, $H_t=4.9$ m, and $H_r=3.4$ m based on \cite{Measurement of Distributed Antenna Systems at 2.4 GHz in a Realistic Subway Tunnel Environment_IEEE Trans. Veh. Technol._2012, Two-Slope Path Loss Model for Curved-Tunnel Environment With Concept of Break Point_IEEE Trans. Intell. Transp._2021}. Representative values $\sigma_\delta\in\{0.1,0.5\}$\footnote{Given train speed $v$ and carrier $f_c$, the Doppler is $f_D\approx (v/c)f_c$. If a fractional residual $\varepsilon$ of $f_D$ remains after compensation, the per-sample residual phase (sampling interval $T_s$) is $ \delta \approx 2\pi\,\varepsilon f_D T_s = 2\pi\,\varepsilon \frac{v}{c} f_c T_s.$ For $v=350\ \mathrm{km/h}$ ($v\!\approx\!97.22$ m/s), the resulting Doppler shifts are $f_D\!\approx\!583$ Hz at $f_c=1.8$ GHz, $f_D\!\approx\!680$ Hz at $f_c=2.1$ GHz, and $f_D\!\approx\!1.13$ kHz at $f_c=3.5$ GHz. With $T_s=1$ ms, a $1\%$ residual ($\varepsilon=0.01$) gives $\delta\!\approx\!0.037$ rad at $1.8$ GHz, $\delta\!\approx\!0.043$ rad at $2.1$ GHz, and $\delta\!\approx\!0.071$ rad at $3.5$ GHz; a $10\%$ residual yields $\delta\!\approx\!0.37$ rad, $0.43$ rad, and $0.71$ rad, respectively.  Hence the representative values $\sigma_\delta\in\{0.1,\,0.5\}$ used in the simulations correspond to small and moderate residual regimes for realistic choices of $(v,f_c,T_s,\varepsilon)$ and thus illustrate sensitivity to residual phase errors.} are used to illustrate the small and moderate residual regimes. The distances $d_h$, ${d_{fi}}$, and ${d_{gi}}$ adhere to the relationships outlined in Section \ref{sec:SysModel}.
	All simulations are statistically averaged over ${10^5}$ independent runs.

	Fig.~\ref{fig_2_PDF} illustrates the simulated PDFs of $G_a$ and  $|G_a|^2$ along with a Gaussian approximation for $G_a$ and a Gamma approximation for $|G_a|^2$ when residual perturbation $\sigma_\delta=0.1$. The Gamma distribution, derived using the parameters in (\ref{eq:gamma parameter1}), is robust aligns closely with the simulated distribution regardless of the value of $N$, while the Gaussian distribution derived using the parameters in (\ref{mu_i,nu_i}) provides a good approximation for large $N$ values but is less so for small $N$ values. 

     \begin{figure*}[t]
    \centering

    % ================= Fig. 3 =================
    \begin{minipage}[t]{0.3\textwidth}
        \centering
        \includegraphics[width=\linewidth]{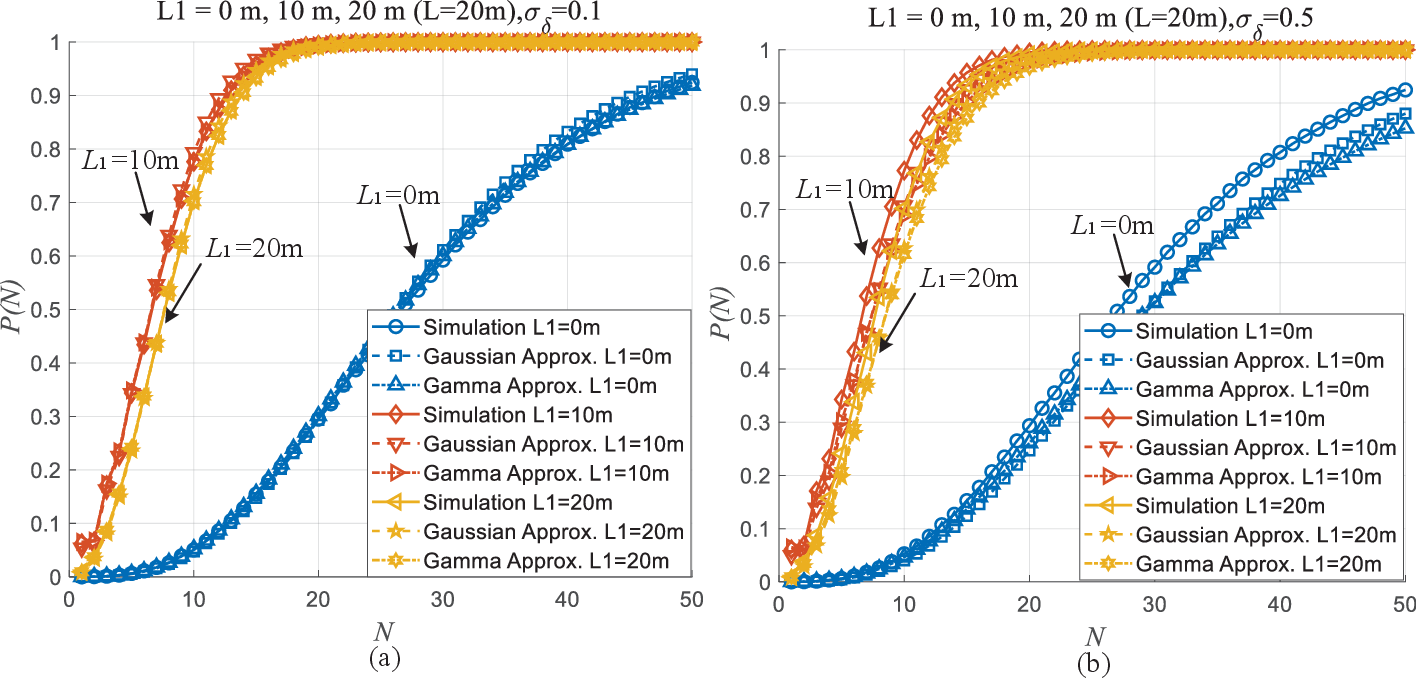}
        \caption{$P(N)$ versus  
                $N$ 
                in the adjustable tag phase case when (a) $\sigma_{\delta}=0.1$ and (b) $\sigma_{\delta}=0.5$.}
        \label{fig_3_probability}
    \end{minipage}\hfill
    %
    % ================= Fig. 4 =================
    \begin{minipage}[t]{0.3\textwidth}
        \centering
        \includegraphics[width=\linewidth]{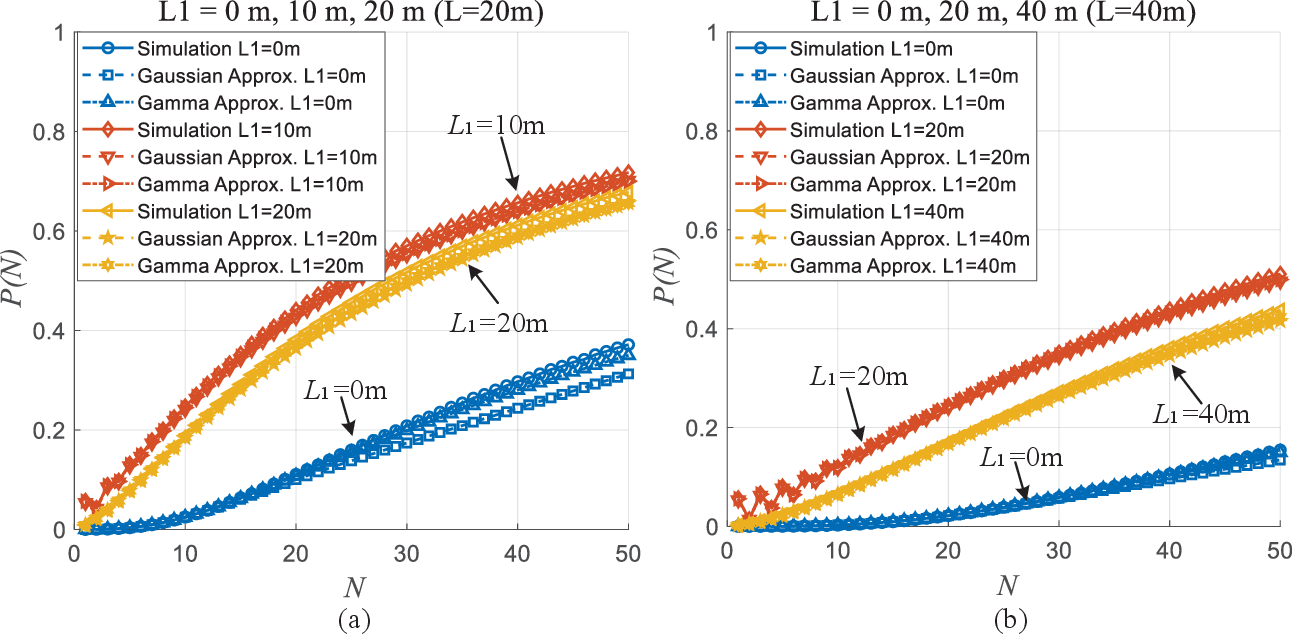}
        \caption{$P(N)$ versus 
            $N$ in the random tag phase case when (a) $L=20$m and (b) $L=40$m.}
        \label{fig_4_probability}
    \end{minipage}\hfill
    %
    % ================= Fig. 5 =================
    \begin{minipage}[t]{0.28\textwidth}
        \centering
        \includegraphics[width=\linewidth]{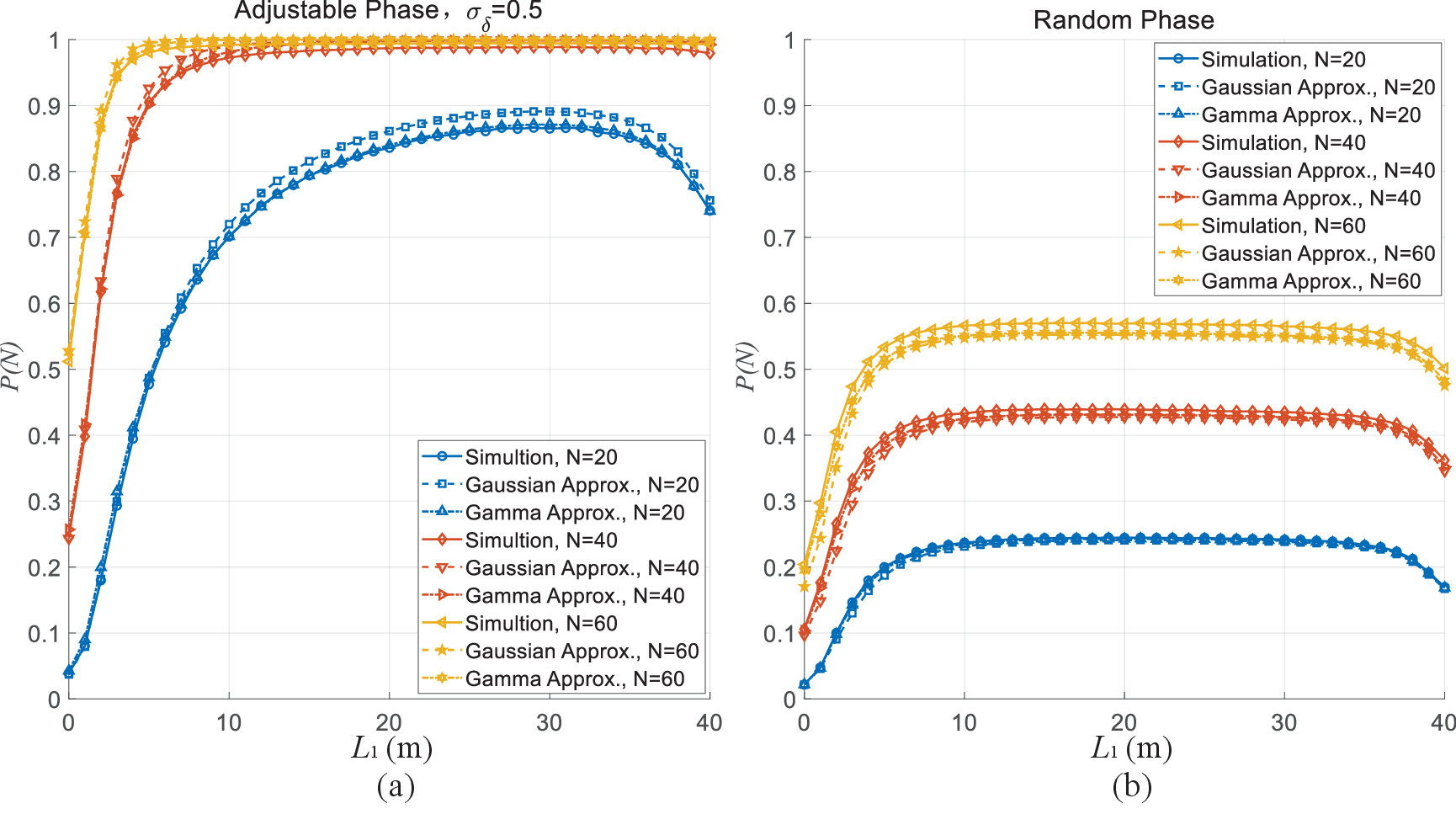}
        \caption{$P(N)$ versus distance $L_1$ for (a) adjustable phase with residual disturbance $\sigma_\delta=0.5$ and (b) random tag phase case when $N=20,40,60$ ($L=40$m).}
        \label{fig_5_probability}
    \end{minipage}

\end{figure*}
	Figs. \ref{fig_3_probability} and  \ref{fig_4_probability} illustrate the probability $P(N)$ as a function of $N$ in the adjustable tag phase case with two different degrees of residual disturbance and random tag phase case, respectively. 
	We can observe that our theoretical derivation results generally coincide with the simulations in all cases, validating the derivations of (\ref{eq:PN_gauss_final}), (\ref{eq:PN_gamma_final}), (\ref{eq:P_gauss_delta}), and (\ref{eq:P_gamma_random}).
	In Fig. \ref{fig_3_probability}, $P(N)$ increases monotonically with $N$ due to the coherent combining of the backscatter components. The growth rate of $P(N)$ is more pronounced when the Rx is closer to the center of the tunnel, providing more balanced link distances and thus stronger constructive aggregation. In contrast, the gain from increasing $N$ in Fig. \ref{fig_4_probability}  is substantially weaker because the backscatter signals add non-coherently. Therefore, the performance gap between adjustable and random phases highlights the importance of tag phase control. 
	Furthermore, $P(N)$ does not increase indefinitely in both cases, that is, $P(N)$ saturates after $N$ exceeds a certain threshold, indicating an upper bound on the effective number of tags. 
    We note that residual perturbations $\delta_i$ reduce the effective coherence and increase the variance of the combined backscatter term in the adjustable case, which explains the small deviation between simulations and the Gaussian/Gamma approximations.  Crucially, the trend of $P(N)$ and the saturation value $N$ are only mildly affected, which reflects that the system retains substantial coherent gain and thus remains robust to moderate residual errors.

	Fig. \ref{fig_5_probability} shows the variation of $P(N)$ with respect to the position of Rx for different numbers of tag when the tunnel length is $40$m. Fig. \ref{fig_5_probability}(a) corresponds to adjustable tag phases with residual disturbance $\sigma_\delta=0.5$, while Fig. \ref{fig_5_probability}(b) corresponds to random tag phases. $P(N)$ exhibits a non-monotonic behavior when tag phases are adjustable. The probability increases as the Rx moves deeper into the tunnel, since the direct link weakens and the relative contribution of backscatter-assisted links grows. It then peaks near the center of the tunnel, where symmetric propagation paths provide stronger constructive combining, and gradually decreases as the Rx approaches the opposite edge, particularly for smaller $N$. This peak becomes higher as $N$ increases, highlighting enhanced coherent aggregation with more tags.
	In contrast, the random phase case shows much flatter curves, because incoherent addition suppresses spatial diversity gains and the probability gain from tags remains limited across different Rx positions.

	\section{Conclusion}\label{sec:conclusion}
	Backscatter technology 
	has the potential 
	to enhance the  receiving signal strength 
	in straight tunnels.   
	This paper studied the channel gain of backscatter-assisted communication under specific tag arrangements in straight tunnel scenarios for high-speed railways. The obtained results showed that phase-adjustable tags can significantly enhance performance, while random tag phases yield limited gains. Moreover, the probability $P(N)$ saturates when the number of tags exceeds a threshold, indicating that further increasing $N$ brings no additional benefit.

	\appendices
	
	\ifCLASSOPTIONcaptionsoff
	\newpage
	\fi

	\vfill

\vspace{-3mm}
	\begin{thebibliography}{00}
		\bibliographystyle{IEEEtran}
		
		\bibitem{Radio Wave Propagation Scene Partitioning for High-Speed Rails_Int J Antenn Propag_2016}  B.~{Ai}, R.~{He}, Z.~{Zhong}, K.~{Guan}, B.~{Chen}, P.~{Liu}, and Y.~{Li}, ``Radio Wave Propagation Scene Partitioning for High-Speed Rails," {\it Int J Antenn Propag.}, vol. 2012, no. 21, pp. 1072-1075, 2016.
				
		\bibitem{3D Non-Stationary Wideband Tunnel Channel Models for 5G High-Speed Train Wireless Communications_IEEE T Intell Transp_2020}  Y.~{Liu}, C. -X.~{Wang}, C. F.~{Lopez}, G.~{Goussetis}, Y.~{Yang},  and G. K.~{Karagiannidis}, ``3D Non-Stationary Wideband Tunnel Channel Models for 5G High-Speed Train Wireless Communications," {\it IEEE T Intell Transp.}, vol. 21, no. 1, pp. 259-272, Jan. 2020.
		
		\bibitem{Intelligent Beam Management Based on Deep Reinforcement Learning_IEEE Trans. Veh. Tech._2024}  Y.~ {Qiao}, Y.~{Niu}, X.~{Zhang}, S.~{Chen}, Z.~{Zhong},  and N.~{Wang}, ``Intelligent Beam Management Based on Deep Reinforcement Learning in High-Speed Railway Scenarios," {\it IEEE Trans. Veh. Tech.}, vol. 73, no. 3, pp. 3917-3931, Mar. 2024.
		
		
		\bibitem{DMRS-based channel estimation for railway communications in tunnel environments_Veh Commun._2021}  C.~{Skiribou}, F.~{Elbahhar}, and R.~{Elassali},  ``DMRS-based Channel Estimation for Railway Communications in Tunnel Environments," {\it Veh Commun.}, pp. 29, Jun. 2021.
		
		
		\bibitem{Activation Distance and Capacity Analysis for Ambient Backscatter Communications with Sensitivity Constraint and Beamforming_China Commun._2023}  Y.~{Mu}, D.~{Fan}, G.~{Wang}, Y.~{Xu} and L.~{Kuang},  ``Activation Distance and Capacity Analysis for Ambient Backscatter Communications with Sensitivity Constraint and Beamforming," {\it China Commun.}, vol. 21, no. 11, pp. 257-266, Nov. 2024.
		
		
		\bibitem{Breaking the Interference and Fading Gridlock in Backscatter Communications_IEEE Commun._2024}  B. Gu, D. Li, H. Ding, G. Wang and C. Tellambura, ``Breaking the Interference and Fading Gridlock in Backscatter Communications: State-of-the-Art, Design Challenges, and Future Directions,'' \emph{IEEE Commun. Surv. Tut.} vol. 27, no. 2, pp. 870-911, Apr. 2025.
		
		
		\bibitem{Channel Estimation for Backscatter Communication Systems_IET Commu_2024} Y.~{Mu}, C.~{Yao}, Y.~{Xu}, G.~{Wang}, M.~{Milošević}, and B.~{Ai}, ``Channel Estimation for Backscatter Communication Systems with Retrodirective Arrays," {\it IET Commun.}, vol. 18, pp. 671-678, 2024.
		
		
		\bibitem{Wavy Signals and Striped Constellations for Backscatter Communications_IEEE Trans. Wireless Commun._2024}  Z.~{Cui}, G.~{Wang}, M.~{Liu}, B.~{Ai}, T. Q. S.~{Quek}, and C.~{Tellambura}, ``Wavy Signals and Striped Constellations for Backscatter Communications: Origins and Solutions," {\it IEEE Trans. Wireless Commun.}, vol. 23. no. 10, pp. 12815-12829, 2024.
		
		\bibitem{Backscatter Aided Wireless Communications on High Speed Rails: Capacity Analysis and Transceiver Design_IEEE J. Sel. Area. Comm._2020}  W.~{Zhao}, G.~{Wang}, B.~{Ai}, J.~{Li}, and C.~{Tellambura},  ``Backscatter Aided Wireless Communications on High Speed Rails: Capacity Analysis and Transceiver Design," {\it IEEE J. Sel. Area. Comm.}, vol. pp, no. 99, pp. 1-1, 2020.
		
		
		
		\bibitem{Backscatter Assisted Wireless Communications_China Commun_2024} Y.~{Mu}, G.~{Wang}, S.~{Atapattu}, X.~{Li}, and B.~{Ai}, ``Backscatter Assisted Wireless Communications for Straight Tunnel Scenario on High-Speed Railways: Signal Strength and Tag Position Analysis,". {\it China Commun.}, to be published, DOI: https://doi.org/ 10.23919/JCC.ja.2023-0634.
		
		\bibitem{Two-Slope Path Loss Model for Curved-Tunnel Environment With Concept of Break Point_IEEE Trans. Intell. Transp._2021}  S. K.~{Kalyankar}, Y. H.~{Lee}, and Y. S.~{Meng},  ``Two-Slope Path Loss Model for Curved-Tunnel Environment With Concept of Break Point," {\it IEEE Trans. Intell. Transp.}, vol. 22, no. 12, pp.  7850-7859, Dec. 2021.
		
		\bibitem{Measurement of Distributed Antenna Systems at 2.4 GHz in a Realistic Subway Tunnel Environment_IEEE Trans. Veh. Technol._2012}  K.~{Guan}, Z.~{Zhong}, J. I.~{Alonso}, and C. ~{Briso-Rodriguez},  ``Measurement of Distributed Antenna Systems at 2.4 GHz in a Realistic Subway Tunnel Environment," {\it IEEE Trans. Veh. Technol.}, vol. 61, no. 2, pp. 834-837, Feb. 2012. 
		
		\bibitem{Wideband Polarimetric Directional Propagation Channel Analysis Inside an Arched Tunnel_IEEE Trans. Antenn. Propag._2009} G. S. Ching, M. Ghoraishi, M. Landmann, N. Lertsirisopon, J. Takada, and T. Imai, ``Wideband Polarimetric Directional Propagation Channel Analysis Inside an Arched Tunnel," {\it IEEE Trans. Antenn. Propag.}, vol. 57, no. 3, pp. 760-767, Mar. 2009.

       

        \bibitem{Transmissive Metasurfaces Assisted Wireless Communications on Railways: Channel Strength Evaluation and Performance Analysis_IEEE Trans. Commun._2023}  J.~ {Lin}, G. ~{Wang}, S. ~{Atapattu}, R. ~{He}, G. ~{Yang}, and C. ~{Tellambura},  ``Transmissive Metasurfaces Assisted Wireless Communications on Railways: Channel Strength Evaluation and Performance Analysis," {\it IEEE Trans. Commun.}, vol. 71. no. 3, pp. 1827-1841, 2023.
		
		\bibitem{Channel Estimation and Pilot Design for Ambient Internet of Things With Frequency Offsets_IEEE Trans. Veh. Tech._2025} Y. Xu, G. Wang, R. Xu, Y. Liu, C. Tellambura and B. Liu, ``Channel Estimation and Pilot Design for Ambient Internet of Things With Frequency Offsets", {\it IEEE Trans. Veh. Tech.},  vol. 74, no. 3, pp. 5125-5129, Mar. 2025.
        
		\bibitem{Doppler mitigation method aided by reconfigurable intelligent surfaces for high-speed channels_WCL_2022} W. Wu, H. Wang, W. Wang, and R. Song, ``Doppler mitigation method aided by reconfigurable intelligent surfaces for high-speed channels", {\it IEEE Wireless Commun. Lett.}, vol. 11, no. 3, pp. 627–631, Mar. 2022.
				
		
		
	\end{thebibliography}
\end{document}